\documentstyle[12pt,epsf]{article}

\begin{document}

\newcommand{\sheptitle}
{SN1a data and the CMB of Modified Curvature at Short and Large Distances}
\newcommand{\shepauthor}
{ Mar Bastero-Gil and  Laura Mersini}

\newcommand{\shepaddress}
{Scuola Normale Superiore and INFN, Piazza dei Cavalieri 7,\\
I-56126 Pisa, Italy}

\newcommand{\shepabstract}
{}

\begin{titlepage}
\begin{flushright}
hep-ph/0107256\\
SNS-PH/01-10\\
\today
\end{flushright}
\vspace{.1in}
\begin{center}
{\large{\bf \sheptitle}}
\bigskip \medskip \\ \shepauthor \\ \mbox{} \\ {\it \shepaddress} \\
\vspace{.5in}

\bigskip \end{center} \setcounter{page}{0}
\shepabstract
\begin{abstract}

The SN1a data, although inconclusive, when combined with other
observations makes a strong case that our universe is presently
dominated by dark energy. We investigate the possibility that large
distance modifications of the curvature of the universe would perhaps
offer an alternative explanation of the observation. Our calculations
indicate that a universe made up of no dark energy but instead, with a
modified curvature at large scales, is not scale-invariant, therefore
quite likely it is ruled out by the CMB observations. The sensitivity
of the CMB spectrum is checked for the whole range of
mode modifications of large or short distance physics. The spectrum
is robust against modifications of 
short-distance physics  and the UV cutoff when:
the initial state is the adiabatic vacuum, and the inflationary
background space is de Sitter. 
\end{abstract}

\vspace*{3cm}

\begin{flushleft}
\hspace*{0.9cm} \begin{tabular}{l} \\ \hline {\small E-mails:
bastero@cibs.sns.it, mersini@cibs.sns.it }
\end{tabular}
\end{flushleft}

\end{titlepage}


\section{Introduction}

Based on the theoretical cosmological models of inflation, the
interpretation of the current astrophysical observations such as the
SN1a data \cite{sn1a}, suggest that our universe contains a large amount
of dark energy \cite{dark}.

However, alternative models, free of dark energy, which may fit in the
allowed range of parameters suggested by observation, are not
excluded.  In this paper we investigate claims to a possibly different
interpretation of the SN1A data, for these alternative
cosmological models: a FRW universe with no dark energy but
with a modified curvature at large enough distances. The hope then is
that either the Friedman 
equation for the expansion is modified, or that the light from SN1a
that reaches us, while passing through these regions of different
curvature, would be deflected, thereby ``appearing'' to have the same
effect as an accelerating universe\footnote{We thank A. Riotto for
bringing this idea to our attention.}$^{,}$\footnote{See
however Ref. \cite{martins} for constraints on models with spatial variations
of the vacuum energy density.}.
 
We examine metric perturbations in this modified background geometry
(traced back at the time of inflation). Metric 
perturbations are responsible for the generation of the large scale
structure and temperature anisotropies of the CMB. The inflaton field
(in 4 dimensions), 
through the Friedmann equation, determines the expansion rate $H$ for
the curvature of the background geometry. The metric perturbations
satisfy a Klein-Gordon scalar field equation, minimally coupled to
gravity \cite{equations}. The scalar field
has a generalized mass squared $\Omega_n(\eta)^2$ that receives the
contributions of two terms: the field frequency squared and the field
coupling to the background curvature term. The coupling of the field to
the curvature results in a modified propagation at long wavelengths
since the curvature of the 
universe is modified at large distances compared to the intermediate
scales. Examples of modified gravity can be found in
\cite{kogan,csaki,gregory,dvali}. Then, the modified propagation of
wavelengths 
of the same scale as the background curvature deviation scale  can be
attributed to a nonlinear dispersed frequency of the field at those
wavelengths, for as long as the generalized mass squared,
$\Omega_n(\eta)^{2}$ in the field equation, remains the same. This
equivalence noticed in \cite{paper1} is very useful for 
calculating the effects of modified large distance curvature in
observations.

Our model consists of a (2-parameter) family of nonlinear dispersion
relations for the generalized frequency of the field, that take account
of the modification of the curvature at large distances.  The family
of dispersion functions is nearly linear for most of the range
$k<M_{P}$, except a nonlinear deviation centered around some low value
of momenta $k_0$. It is this deviation bump that reflects the
modifications of the generalized frequency of the field at low momenta
$k_0$ due to the modification of the curvature at large distances
$k^{-1}_{0}$.  The dispersion function introduced in Sect. 2, although
nonlinear in the transplanckian regime, it is nevertheless a smooth
function there, asymptotically approaching a constant value at
time-infinity, thus having a well defined initial vacuum state
\cite{laura}. The analytical calculation of the CMB spectrum is based
on the Bogoliubov 
coefficient method. The details of the exact solutions for this class of
dispersion functions \cite{laura} are given in the Appendix. 

In Sect. 3 we check the sensitivity of CMB
spectrum to the bump parameters $k_0$ and  $B$ (scale location and
amplitude) that control the deviation
behavior  from a linear frequency dispersion
at low values of the momenta; i.e., the allowed range of curvature
modifications at very large or very short distances that may agree with observation. 
 We use
CMBFAST in Sect. 3 to plot the spectrum, by replacing the standard
primordial power spectrum $\delta_H^0 (k)$ with that derived
analytically in Sect. 2 for the model considered. We
comment and summarize the results in Section 4. It is shown that the
CMB spectrum is sensitive only to the choice of the initial vacuum
state and the departure from linearity in the low momenta
regime. However, for an adiabatic initial vacuum state, the CMB spectrum
of a de Sitter expansion  does not depend in the details of
nonlinearity in the transplanckian regime
\cite{brand,niemeyer,akim,paper1,starobinsky,easther} .
 
\section{The Model}
The generalized Friedmann-Lemaitre-Robertson-Walker
(FLRW) line-element  in the presence of scalar and tensor
perturbations, takes the form \cite{FLRW1}
\begin{eqnarray}
ds^2&=& a^2(\eta) \left\{ -d\eta^2 + \left[ \delta_{ij} + h(\eta,{\bf
n})Q \delta_{ij} \right. \right.  \nonumber \\ & &\left. \left. + h_l
(\eta,{\bf n}) \frac{Q_{ij}}{n^2} + h_{gw}(\eta,{\bf n}) Q_{ij}
\right] d x^i d x^j \right\} \,,
\label{frw}
\end{eqnarray}
where $\eta$ is the conformal time and $a(\eta)$ the scale factor. The
dimensionless quantity ${\bf n}$ is the comoving wavevector, related
to the physical vector ${\bf k}$ by ${\bf k}= {\bf n}/a(\eta)$. 
The function $(h,\,h_l)$ and $h_{gw}$ represent the {\em scalar and
tensor} perturbations respectively.

The power spectrum of the perturbations can be computed once we solve
the equations in the scalar and tensor sector. The equation for the
metric perturbations corresponds to a Klein-Gordon equation of a
minimally coupled scalar field, $\mu_n$, in a time dependent
background\footnote{We refer the reader 
for the details of the procedure to Refs. \cite{mode1} and
related references \cite{equations}.}
\begin{equation}  
\mu_n^{\prime \prime} + \Omega_n(\eta)^2 \mu_n=0 \,,
\label{kg}
\end{equation}
where the prime denotes derivative with respect to conformal time $\eta$, and 
the generalised comoving frequency is\footnote{Note that from here on
we use the symbol $a$ instead of $a(\eta)$ for the scale factor.}

\begin{equation}
\Omega_n(\eta)^2 = n^2 - \frac{a^{\prime \prime}}{a}= a^2 k^2 -\frac{a^{\prime
\prime}}{a}  \,.
\label{omegan}
\end{equation}

The dynamics of the scale factor is determined by the
evolution of the background inflaton field $\phi$, with potential
$V(\phi)$, and the Friedmann equation. 
There are mechanisms that may produce different scale factors by
modifying gravity at large (e.g. \cite{kogan,csaki,gregory,dvali}) or
short distances (\cite {akim}).The present large distance modification
scales can be traced back 
in time and would correspond to deviations in the primordial scale
factor and spectrum.  We can denote
this ``distance-dependent'' scale-factor by ${\cal A}$. 
The coupling of the field to this background curvature results in a
modified propagation of the field at 
long wavelengths. Therefore, modifications of the scale factor or
curvature ($\mathcal{A}$) of the universe at large scales  
can be attributed to a dispersed effective frequency $(n_{eff})$, 
such that the generalized comoving
frequency  Eq. (\ref{omegan}) remains the same, in the following
manner\begin{equation}
\Omega_n(\eta)^2=n^2 -\frac{\mathcal{A}^ {\prime
\prime}}{\mathcal{A}}=n_{eff}^2 -\,\frac{a^{\prime \prime}}{a}\,. 
\label{omeganef}
\end{equation}
$n_{eff}$ denotes the 
dispersed comoving frequency of the field due to absorbing the
modification terms to the curvature, ${\cal A}''/{\cal A}$. Therefore, the 
dispersion function for the generalized
frequency results from the modified curvature at very large
 and very short distances. It deviates from linearity at small 
momentum $k$ and asymptotically approaches a constant value in the
transplanckian regime.

\begin{figure}[t]
\begin{tabular}{cc} 
\epsfysize=8cm \epsfxsize=8cm \hfil \epsfbox{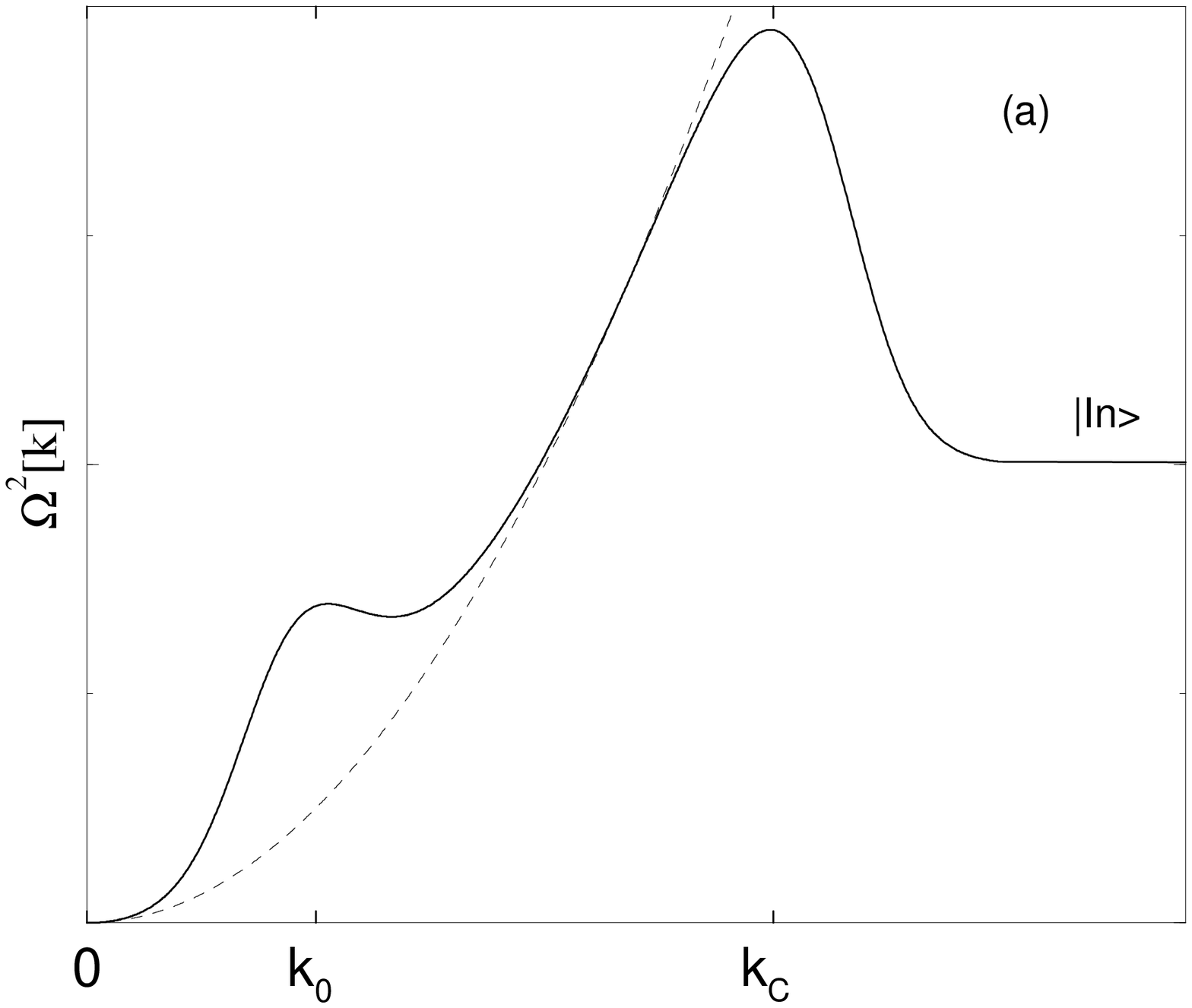} \hfil
&
\epsfysize=8cm \epsfxsize=8cm \hfil \epsfbox{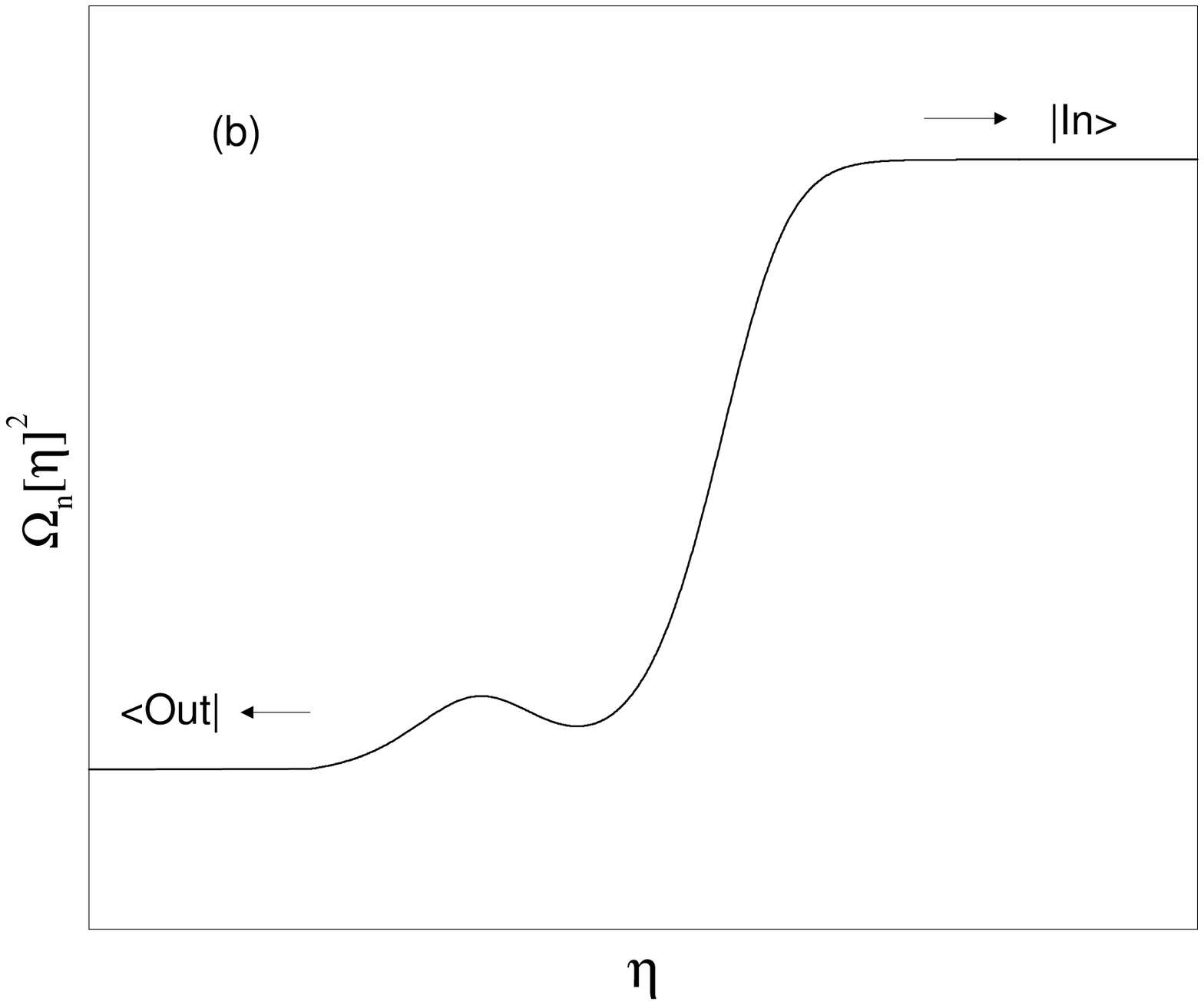} \hfil
\end{tabular}
\caption{{\footnotesize Shown is one of the members of the family of
dispersion relations $\Omega_n(\eta)^2$ as a function of: (a)$\Omega_k^{2} =a^2 F[k]^2$ as a function of the physical
momentum $x$; (b) conformal time $\eta$. The variable $x=k/k_C$ has
been shifted by 1 
such that the regime of linear dispersion relation corresponds to
small positive 
values of momenta, i.e., $x\ll1$.} }
\label{fig1}
\end{figure}

The dispersion relation for the generalized comoving frequency
$\Omega_n(\eta)$ is simply\footnote{From here on, we absorb the term 
$a^{\prime \prime}/a$ of Eq. (\ref{omeganef}) into the definition
of the dispersion function $F[k]$.} \cite{laura}: 
$\Omega_n(\eta)= a(\eta) F[n/a(\eta)]$. 
The 2-parameter family of dispersion functions $F[k]$ of our model (see
Fig. \ref{fig1}) is:
\begin{eqnarray} 
F[k]^2 &=& (k^2-k_1^2) V_0(x,x_0) + k^2~V_1 (x-x_0) + k_1^2 \,,
\label{omegak} \\
V_0(x,x_0) &=& \frac{C}{1+e^x} + \frac{ E~e^x}{(1+e^x)(1+e^{(x-x_0)})} \,, \\
V_1(x-x_0) &=& -B \frac{e^x}{(1+e^{(x-x_0)})^2} \,, 
\label{omegak2}
\end{eqnarray}
where the dimensionless wavevector is $x=k/k_C$, $k_C=M_P$ is the cutoff
scale, $k_0 \ll k_C$, (i.e. $x_0 \ll \ 1)$ is the  
value at which we deviate from linearity at low momentum,  the
deviation amplitude is controlled by the parameter $B$, and the constant parameter
$k_1 < k_C$  is the asymptotic value of the frequency in the transplanckian
regime ($k \rightarrow \infty$). $C,E,B,x_0$ are dimensionless parameters. 

As already discussed in Refs. \cite{brand,niemeyer,akim},
Eq. (\ref{kg}) represents 
particle production in a time-dependent background
\cite{partcreat,birrell}. We will follow the 
method of Bogoliubov transformation to calculate the spectrum.  
The frequency $\Omega_n(\eta)^2$ (which is the same as a `time-dependent mass
squared` term) goes asymptotically to constant values at late 
and early times. Therefore the initial and final vacuum states are well
defined. At early times the wavefunction should behave as a plane wave:
\begin{equation}
\mu_n \rightarrow_{\eta \rightarrow - \infty}
\frac{1}{\sqrt{2\Omega_n^{in}}} e^{-i \Omega^{in}_n \eta} \,.
\end{equation}
But at late times one has a squeezed state due to the curved background
that mixes positive and negative frequencies. The evolution of the
mode function $\mu_n$ at late times fixes the Bogoliubov coefficients
$\alpha_n$ and $\beta_n$,
\begin{equation}
\mu_n \rightarrow_{\eta \rightarrow +\infty} \frac{\alpha_n}{\sqrt{2
\Omega^{out}_n}} e^{-i\Omega^{out}_n \eta} + \frac{\beta_n}{\sqrt{2
\Omega^{out}_n}} e^{+i \Omega^{out}_n \eta}\,.
\end{equation}
with the normalization condition:
\begin{equation}
|\alpha_n|^2 - |\beta_n|^2 =1\,.
\end{equation}
In the above expressions, $\Omega^{in}_n$ and $\Omega^{out}_n$ denote
the asymptotic values of $\Omega_n(\eta)$ when $\eta \rightarrow \mp
\infty$. 

Details of the exact solution for Eq. (\ref{kg}) with the dispersed frequency
given by Eqs. (\ref{omegak}-\ref{omegak2}) are given in the Appendix.
The final expression for the Bogoliubov
coefficient $|\beta_n|^2$ is:  
\begin{equation}
|\beta_n|^2 = \frac{ \sinh^2(2 \pi \hat{\Omega}_-) + \Gamma(k_0,B)}
{\sinh^2(2 \pi \hat{\Omega}_+) - \sinh^2(2 \pi \hat{\Omega}_-)} \,,
\label{betak}
\end{equation}
where  $\hat{\Omega}_i=\Omega_i/n$, $\hat{\Omega}_{\pm}=(\hat{\Omega}^{out} \pm
\hat{\Omega}^{in})/2$, and the deviation
function $\Gamma(k_0,B)$ that contains the departure from
thermality in the spectrum is  
\begin{equation} 
\Gamma(k_0,B)= \cosh^2 (\frac{\pi}{2} \sqrt{ 4 B e^{-x_0} - 1})\,.
\end{equation}
When $B=0$, for $\Omega^{in} > \Omega^{out}$, then it is clear from Eq. (\ref{betak}) that the   
spectrum of created particles is nearly thermal to high accuracy,
\begin{equation}
\left|\frac{\beta_n}{\alpha_n} \right|^2 \simeq e^{-4 \pi \sqrt{C}} \,.
\end{equation} 
The function $\Gamma(k_0,B)$ represents the $deviation$ of the spectrum from thermal
behavior due to the non-linearities at low momentum. Therefore, the
amplitude of the power spectrum, $\delta_H(k)$, will be modified by
$\Gamma(k_0,B)$ due to the non-linear dispersion function introduced
at around $x_0 <1$.  

In de Sitter space, the Bogoliubov coefficients would not depend on $k$
except their dependence in the bump parameters through the deviation
function $\Gamma(k_0, B)$. This function represents the departure from
thermality in the particle creation number, $|\beta_n|^2$ and it confirms
B. L. Hu idea \cite {blhu} that near thermal radiance can be
characterized by departure from exponential scaling. 
It is straightforward to derive the CMB power spectrum, $P(n)$,
analytically from (the exact solution for) the Bogoliubov coefficients
$\alpha_n, \beta_n$ \cite{easther} 
\begin{equation}
P(n) = \frac{n^3}{2\pi^2}|\frac{\mu_n}{a}|^2 \simeq |\beta_n + \alpha_n|^2 \,.
\end{equation}
The deviation of the spectrum from scale-invariance in this class of
models depends on the parameters of large-distance curvature
modifications, namely: the scale of modified long wavelength modes, $k_0^{-1}$,
and the deviation amplitude $B$.  

The expression for the Bogoliubov coefficient and Eq. (13) indicate
that: for a well-defined 
initial vacuum state\footnote{The field is in an
initial Bunch-Davies vacuum.}, the spectrum is insensitive to the 
nonlinear dispersion relation in the transplanckian regime
(modifications of short-distance physics). The unusual CMB spectrum
plotted in the next Section with CMBFAST, demonstrates that
modifications of the large scale curvature of the universe 
produce a tilt due to the departure from scale-invariance, and
therefore conflict with the observed CMBR 
spectrum. In general the tilt is enhanced for modifications at
superhorizon scales ($k_0\leq H_0$) because it is the low energy modes
that dominate the spectrum in the Bogoliubov coefficient. 
Although departure from scale invariance is smaller at 
the last scattering horizon scale, $H_{LS}$, the range of deviation
parameters is constrained by the amplitude of the first peak. The
deviation introduced to the spectral index, $n_s$ 
from higher energy modes (wavelengths shorter than the last
scattering horizon $k_0 >H_{LS}$) becomes negligible because high energy modes do
not contribute significantly to the spectrum. However, the shorter
wavelengths would correspond to the intermediate FRW regime rather
than the large distance scales, a regime which is scrutinized by
direct observation.

\section{CMB Spectrum}

Recent Boomerang and MAXIMA-1 CMB experiments \cite{cmb1,cmb2} have, to high
accuracy,  constrained the cosmological parameters, derived from the
family of inflationary adiabatic models, to: total energy density $\Omega_{tot}=0.90 \pm 0.15$
and spectral index $n_s=0.99 \pm 0.09$ at a 95\% confidence level \cite{cmb3}. 
The current data favors a universe with dark energy density
$\Omega_{\Lambda}=0.7$.

In this part, we explore the cosmological consequences of the
alternative model that was given in Section 2 (Fig. 1). CMB is the
most difficult test of precision cosmology that 
these models should pass. This model contains no dark energy,
$\Omega_\Lambda=0$, however it 
describes  a universe which at large distances has a modified
curvature from the metric of the FRW universe at intermediate scale. 
In Fig.2 we show the CMB power spectra obtained for different
representative values of the deviation parameters $k_0$ and $B$ in the
dispersion function Eqs. (\ref{omegak}-\ref{omegak2}). The conventional parameters
that go in the input of CMBFAST are: $(\Omega_{tot}, \Omega_b,
\Omega_c ,\Omega_{\Lambda}$), which stand for total energy density,
baryonic, cold dark 
matter and the cosmological constant energy density respectively;  
and $n_s$ which is the scalar spectral index. We modified the power
spectrum amplitude $\delta_H^0(k)$ in the POWERSFLAT subroutine of
CMBFAST, in order to contain the deviation from the thermal spectrum 
(for the exact calculation reported in Section 2). 
The modified perturbation amplitude $\delta_H^{2}(k)$ is expressed
in terms of $\delta_H^{0},\, k_0,\, B$, where
$\delta_H^0$ is the unmodified amplitude of the
scale-invariant power spectrum, 
$k_0$ corresponds to the location-scale where the curvature is
modified, and B measures the  amplitude of deviation in the curvature at
scale $k_0$.

The values of the conventional parameters were taken to be (1,0.03,
0.97,0) for all the deviation plots ($II-V$), but the deviation parameters in the 4 plots
below in Fig. 2 are in respective order: 

\begin{tabular}{lll}
I &(solid line):&         ($k_0 = 0$, $B=0$, $\Omega_\Lambda=0.7$) \\
II &(long-dashed line):&  ($k_0 = 10^{-6}$ hMpc$^{-1}$, $B=2$)\\
III &(dashed line):&      ($k_0 = 0.05$ hMpc$^{-1}$, $B=2$)\\
IV &(dot-dashed line ):&  ($k_0 = 5$ hMpc$^{-1}$, $B=2$) \\
V &(dotted line ):     &  ($k_0 = 0.05$ hMpc$^{-1}$, $B=2.5$)
\end{tabular}

All plots were normalized to COBE. Shown for comparison is also plot
$I$ corresponding to the conventional CMB spectrum with
$\Omega_{\Lambda} = 0.7$. 
As it can be seen from the plots in Fig. 2, there are distinct
features 
of the CMB spectra corresponding to the dispersion function in
comparison to the standard spectrum obtained for ($\Lambda$)CDM models. 

There is an overall tilt produced in the spectrum which signals
departure from the scale invariance. This tilt is a function of the
amplitude and scale of the modification, $k_0,B$, introduced in Sect.2
(Eq.\ref{betak}), such that it increases 
for low values  of the deviation momentum scale $k_0$ and large 
deviation amplitude $B$. Let us consider the 3 regimes into which the
curvature modifications can be introduced: 

(1) Modifications at superHubble scales ($k_0 < H_0 $). The departure
from scale invariance is the strongest because the low energy modes
dominate the 
spectrum, (II) in Fig. 2. Models predicting curvature modifications
in regime (1) quite likely are ruled out due to a {\em strongly tilted}
spectrum.

\begin{figure}[t]
\epsfxsize=10cm \epsfxsize=10cm \hfil \epsfbox{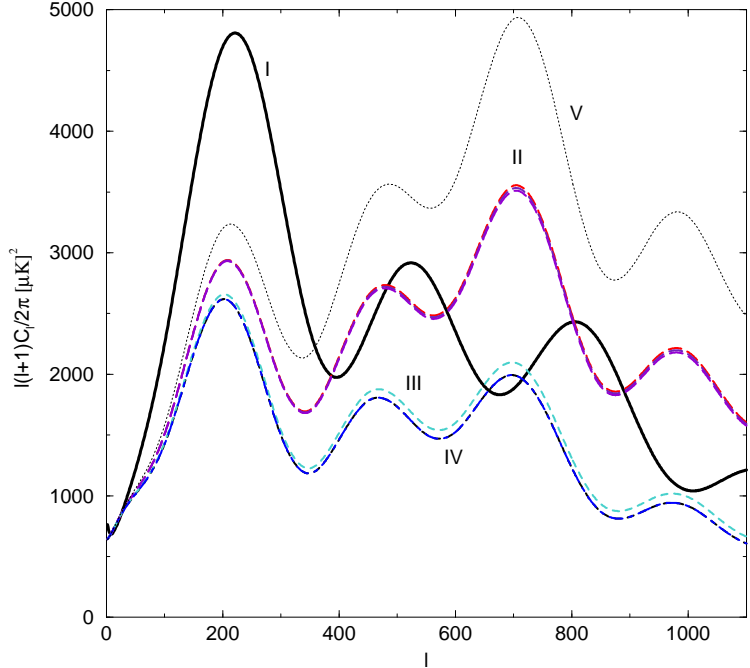} \hfil
\caption{{\footnotesize The CMB spectrum corresponding to our model
for three different representative values of the deviation parameters
$k_0$ and fixed deviation amplitude $B=2$, $II-IV$. Shown is also the
case of a larger amplitude $B=2.5$ at scale $k_0\approx H_{LS}$ (V).
The CMB plots were obtain using CMBFAST and they 
were normalised to COBE. } }
\label{fig2}
\end{figure}

(2) Modifications in the distance range between the current
horizon $H_0$ and last scattering horizon scale $H_{LS}$ ($H_0 \leq
k_0 \leq H_{LS}$). For this range, the tilt is less pronounced then in
regime (1). The main constraint comes from the tilt and
it tightly   
limits the amplitude of deviation in the first peak. For modifications
around the last scattering horizon scale, $k_0 \approx H_{LS}$ the
departure from scale invariance is vanishing, therefore the
constraints are relaxed. However, even in this case the parameter B is
tightly constrained  to deviation by less than 10\%,    
in order for the amplitude of the first peak $A_1$ to be in the
allowed range of 4500-5500 $\mu K^2$ 
\cite{cmb1,cmb2}.  
In Fig. \ref{fig2} we show the CMB spectra for these tuned values of
$k_0,\,B$; for comparison we also plotted the CMB for a  value of
$B=2.5$, which is outside the allowed range. 

(3) Modifications at distances shorter than the last scattering
horizon ($k_0>H_{LS}$). As we approach higher energy modes, the
effect of the modification in the tilt of the spectrum is suppressed,
therefore the departure from the conventional spectrum is
vanishing. Nevertheless, these length scales do not correspond to
large 
distances anymore, instead they are in the intermediate regime of FRW
Universe. Thus the possibility of curvature modifications at such
scales (galactic and intergalactic) is ruled out by direct observation
up to very short distances (less than 1 mm). Clearly, there is no
tilt or departure from the conventional CMB produced in the limit of
modifications of very short distance physics, (very high momenta $k_0
\rightarrow \infty$, i.e., transplanckian regime).    


The claim of the model was to ``offer an alternative explanation'' to
the SN1a data, namely: either the conventional Friedmann equation is
modified or; the light of the SN1a passing through  regions 
of modified curvature would get deflected, and therefore when received
by us would appear as if indicating an accelerating universe.  
Although this alternative approach to the SN1a data might be
theoretically appealing, we conclude that the CMB data tightly
constrains it and makes it unlikely to bear any resemblance to
reality. The method used in this work, can also be adopted to check if
higher dimensional models that predict modified gravity at large
scales and modified equations for the perturbations
\footnote{These models naturally modify the curvature
around horizon and Planck length-scales due to the higher dimensional
gravity effects that switch on at very large or very short distances, 
but nevertheless with contributions from higher graviton excitations
suppressed \cite{rattazzi}.} \cite{kogan,csaki,gregory,dvali} satisfy
the CMB constraints.   

\section{Summary}

In this work we investigated claims that a modified large-distance
curvature may offer an alternative 
explanation for the SN1a data. To check these claims, we studied the
sensitivity of CMB 
spectrum to the whole range of modes, $0 \leq k \leq \infty $, when
short and large distance regimes are modified.

In \cite{paper1} it was noticed that a modified curvature of the
universe at large distance  (when traced back at the time of 
inflation\footnote{It should be noted that in the case of 
higher dimensional multigravity
\cite{kogan,csaki,gregory,dvali,rattazzi}, 
it is not clear how the metric perturbation equations are
modified. })  gives rise to a dispersed frequency for the 
cosmic perturbations. The field is minimally coupled to the curvature
thus its propagation feels the modifications in the background
geometry. We adopted the method of Ref. \cite{paper1} in order to
find out the effects of curvature deviations on the current
astrophysical observables. 

The role of a modified curvature of the universe at large distances on
the inflationary metric perturbations was analytically described by a
family of dispersion relations. The modification modulates the
generalized frequencies of the inflationary perturbation modes at
small values of the momenta $k$ by departing from linearity around
some certain small momenta $k=k_0$ ($k_0 \leq M_{P}$) with a
deviation amplitude $B$. The nonlinear feature of the dispersion
relations, at small momenta $k_0$ and in the transplanckian regime,
tracks the curvature deviations at large and short distances, from the
conventional FRW universe of intermediate scales. One of the
parameters ($k_1^2 < k_C^2$), in this class of dispersion
functions was constrained in order to satisfy the Starobinsky bound
for negligible backreaction \cite{starobinsky}.

The analytical expression for the CMBR spectrum (Sect. 2), as well as
the CMBFAST plots of this class of models, deviate from the {\em black
body scale invariant spectrum}. The deviation function 
$\Gamma(k_0,B)$,  given in Sect. 2 and the Appendix,
 which measures departure from the scale-invariant
spectrum (deviation from thermality in the the Bogoliubov coefficient),
depends on two free parameters, the scale $k_0$  and the amplitude of the
curvature modifications $B$. The tilt produced in the spectrum due to
$\Gamma(k_0,B)$ is present for 
all modification scales $k_0 \leq M_{P}$ (these values of the physical
momenta correspond to the time of inflation). 
The tilt is less pronounced for scale 
modifications corresponding to length scales less or equal to the
horizon of the last scattering surface, and in this case, the main
constraint comes from the 
modifications to the amplitude of the first acoustic peak and the fact
that curvature modifications in the intermediate FRW Universe scales
are under direct observation. It remains interesting to answer why the
only curvature modifications that for a small range of $k_0$ and $B$ can
reconcile with the conventional 
CMB spectrum are allowed only around the last scattering $H_{LS}$
scales.  

The scale and amplitude of the deviations from the conventional
spectrum, are severely constrained from the observed CMB spectrum to
be within $10\%$ of the scale and amplitude of the first
peak. Although it is counterintuitive, since large distance would
correspond to low energy theories, our results indicate that any
modifications in the large scale curvature of the universe, is tightly
constrained from CMB data, to a very small range of deviations from
the curvature of the intermediate FRW universe. Perhaps, there is a
natural way that would explain such a universe with an FRW spacetime
at intermediate and very large distances but with small curvature
deviations around $H_{LS}$, without 
the need to appeal to fine-tuning. But if not, then theoretical
cosmological models would have to account for the negative pressure
dark energy of the universe.

The analysis of the sensitivity of the CMB spectrum for the whole
range of modes in a de Sitter background space, with modifications in the short
and large distance physics, reveal: the spectrum is insensitive
to the details of {\em short-distance physics
and the cuttoff scale} $k_C$ (the transplanckian
regime) only for an {\em initial adiabatic vacuum state};
the scale-invariance of the spectrum and the amplitude of the first
acoustic peak {\em are very sensitive to modifications of large
distance physics} (low momentum modes); {\em the spectrum is also
highly sensitive to the choice of the initial 
conditions}\footnote {It has been argued by many authors \cite {paper1,
starobinsky, niemeyer} that the 
adiabatic vacuum  is the right
choice for the initial conditions. Even 
for the same dispersion model, a different choice for the initial
vacuum state will clearly result in a different particle spectrum,
therefore one has to be careful to distinguish if the features
observed in the CMB spectrum are signatures 
of new physics or only of the choice of initial conditions.}. In our class
of dispersion models, the initial vacuum state is well-defined since
the background ($\Omega_n(\eta)^{2}$) goes asymptotically flat at early
times  (Bunch-Davis vacuum \cite{initial}). The CMB spectrum for this
class of models is indeed insensitive to short distance
modifications, as it can be checked by taking the limit when large
scale modification parameters ${k_0,B}$ go to zero, in  which case the
conventional scale-invariant spectrum is recovered. Therefore
all the features observed in Fig. \ref{fig2}, are due to large-scale curvature
modifications only. 


{\bf Acknowledgment:} We are very grateful to S.Dodelson for his help
with the CMBFAST. We want to thank  A. Riotto, R. Kolb,
L. Parker, A. Kempf, P. Frampton, G. Siegl, I. Kogan, for beneficial
discussions and 
comments. We also would like to thank P. Kanti for useful discussions
in the early stages of this work.  
We acknowledge Lloyd Knox for making CMBFAST program available.

\section{Appendix}

The family of dispersion functions we used in Sect. 2  to model the
deviation of the 
curvature at large and short distances is given by:
\begin{eqnarray} 
F[k]^2 &=& (k^2-k_1^2) V_0(x,x_0) + k^2~V_1(x-x_0) + k_1^2 \,, \label{omegaka}
\\
%
%
V_0(x,x_0) &=& \frac{C}{1+e^x} + \frac{ E~e^x}{(1+e^x)(1+e^{(x-x_0)})}  \,, \\
V_1(x-x_0) &=& -B\frac{e^{x}}{(1+e^{(x-x_0)})^2} \,, 
\end{eqnarray}
where\footnote{The momentum $k$ has been shifted by $k_C$ such that
$\Omega_n \approx x$ for small positive values, $x\ll1$.}  $x=k/k_C$,
$x_0=k_0/k_C$;  $k_C=M_P$ is the cutoff scale, $k_0 
\ll k_C$ ($x_0 \ll 1$) is the   
value at which we deviate from linearity at low momentum, 
and the amplitude of the ``bump/deviation'' is controlled by $B$ (see
Fig. \ref{fig1}). The parameter $k_1=n_1/a(\eta)$ gives  the
asymptotic constant value at initial time for the frequency in the
limit ($k \gg k_C$), i.e., in the transplanckian  regime. 
On the other hand, in order to ensure the linear
behavior at very low values of the momenta, $x \ll 1$, we impose the
following constraints for any value of the deviation parameters $x_0$
and B:
\begin{eqnarray}
V_0(x\ll 1,x_0) &\simeq& 1\,, \;\;\;\;\; V_0(x\ll 1,x_0)^{\prime
\prime} \simeq 0\,, \\
V_1(x\ll 1,x_0) &\simeq& 0\,, 
%
\end{eqnarray}
where prime denotes derivative with respect to the physical momentum $k$. 
The generalised comoving frequency $\Omega_n(\eta)$ is then given by:
\begin{eqnarray}
\Omega_n(\eta)^2&=&a(\eta)^2 F[n/a(\eta)]^2 \nonumber \\
&= & \!\!\!\! (n^2-n_1^2) \left[ \frac{C}{1+e^x} +  \frac{E~
e^x}{(1+e^x) (1+e^{(x-x_0)})} \right] 
- n^2 \left[\frac{B~ e^{x}}{(1+e^{(x-x_0)})^2} \right] + n_1^2 \,,
\end{eqnarray}
with $n=a(\eta) k$, $a(\eta)=-|\eta_C|/\eta$ during de Sitter inflation
($|\eta_C|=1/H(\eta_C)$), and 
\begin{equation}
x=\frac{k}{k_C}= - \frac{n}{|\eta_C| k_C} \eta \,.
\end{equation}
The generalised frequency $\Omega_n(\eta)$ goes to constant values at
$\eta \rightarrow 
\pm \infty$, such that:  
\begin{eqnarray}
\Omega_n(\eta) &\rightarrow_{\eta\rightarrow -\infty}& \Omega_n^{in}=
n_1 \,, \\
\Omega_n(\eta) &\rightarrow_{\eta\rightarrow +\infty}& \Omega_n^{out}=
\sqrt{ n_1^2 + C (n^2-n_1^2)}\,,
\end{eqnarray}
with $\Omega_n^{out}\simeq \sqrt{C} n$ when $n \gg n_1$.

Under the change of variables $\eta \rightarrow u= exp( d n \eta)$, where
$d=1/(|\eta_C| k_C)$, the
scalar wave equation (\ref{kg}) for the mode $\mu_n$ becomes:
\begin{equation}
\left[ \partial^2_u + \frac{1}{u} \partial_u+ V(u) \right] \mu_n =0
\,,
\label{modeva}
\end{equation}
where:
\begin{equation}
V(u)=\hat{D}+ \frac{ \hat{C}}{u (1+u)}+ \frac{\hat{E}}{u (u+1)
(\gamma_0+u)} - \frac{\hat{B}}{u 
(u+\gamma_0)^2} \,,
\label{potx} 
\end{equation}
and, 
\begin{eqnarray*} 
&  &\hat{C} =C [(n^2-n^2_1)/(n~d)^2]\,,\;\;\; \hat{E}= E
[(n^2-n^2_1)/(n~d)^2] \,, 
\\
& &\hat{D}= (n_1/n~d)^2 \,, \\
& & \hat{B} = B/d^2 \,,
\end{eqnarray*}
and $\gamma_0=exp(-k_0/k_C)$. Eq. (\ref{modeva}) is exactly solvable
in terms of the Riemann generalised hypergeometric 
functions \cite{abramowitz} with the constraint
$\hat{E}=\hat{C}/(1 - \gamma_0)$,
\begin{equation}
\mu_n \propto P \left( \begin{array}{cccc}
0 & \infty &-\gamma_0& \\
a & c &b& u \\
a^*& c^*& b^* & \end{array} \right)  \,. 
\end{equation}

As explained in Section 2, because of the asymptotic behavior of
$\Omega_n(\eta)$, the initial and final vacua are well defined and the mode
functions $\mu_n$ 
behave as plane waves in the asymptotic limits $\eta \rightarrow  \mp \infty$. 
The exact solution which matches this asymptotic behavior is then given by:
\begin{eqnarray}
\mu^{in}(\eta)= N^{in} 
(u)^{a} (u+\gamma_0)^b ~ _2F_1 [a+b+ c,a+b+c^*,1+a-a^*,
-\frac{u}{\gamma_0}] \,, 
\end{eqnarray} 
where $N^{in}$ is a normalization constant, and 
\begin{eqnarray}
a&=& -i \hat{\Omega}^{in}=-i \sqrt{\hat{D}} \,,\\
b&=& \frac{1}{2} \left( 1 + \sqrt{ 1 - 4 \hat{B} e^{-x_0}} \right) \,, \\
c&=&  -i \hat{\Omega}^{out} =-i \sqrt{\hat{D} + \hat{C}}\,. 
\end{eqnarray}
At late times the
solution becomes a squeezed state by mixing of positive and negative
frequencies:
\begin{eqnarray}
\mu^{out}_n(\eta)  \!\!\!\!\!\! &=& N^{out} ~ 
(u)^{a} (u+\gamma_0)^b \times \nonumber \\ 
& & \!\!\!\!\!\! \left(  \frac{\Gamma(1+a-a^*)
\Gamma(c^*-c)}{\Gamma(a+b+c^*) \Gamma(1-a^*-b-c)}  
~ _2F_1 [a+b+c,a^*+b+c,1+c-c^*,
-\frac{\gamma_0}{u}] \right. \nonumber \\
& + & \!\!\!\!\!\! \left. 
\frac{\Gamma(1+a-a^*) \Gamma(c^*-c)}{\Gamma(a+b+c) \Gamma(1-a^*-b-c^*)} 
~ _2F_1 [a+b+c^*,a^*+b+c^*,1+c^*-c,-\frac{\gamma_0}{u}] \right) \,,
\nonumber \\
 \mu^{out}_n & \rightarrow_{\eta \rightarrow +\infty}& \frac{\alpha_n}{\sqrt{2
\Omega^{out}_n}} e^{-i\Omega^{out}_n \eta} + \frac{\beta_n}{\sqrt{2
\Omega^{out}_n}} e^{+i \Omega^{out}_n \eta} \,,
\end{eqnarray}
with $|\beta_n|^2$ being the Bogoliubov coefficient equal to the
particle creation number per mode $n$ and $\hat \Omega_i = \Omega_i
/n$.  Using the linear transformation 
properties of hypergeometric functions \cite{abramowitz}, 
we find that
\begin{equation}
\left|\frac{\beta_n}{\alpha_n}\right|^2 = \frac{ \sinh^2(2 \pi
\hat{\Omega}_-) + \Gamma(k_0,\hat{B})} 
{\sinh^2(2 \pi \hat{\Omega}_+)+\Gamma(k_0,\hat{B})} \,,
\label{betaka}
\end{equation}
where,
\begin{eqnarray}
\hat{\Omega}_{\pm}= \frac{\hat{\Omega}^{out} \pm
\hat{\Omega}^{in}}{2} \,,\;\;\;\; \hat{\Omega}^(i)= \frac{\Omega^i}{nb}\,,
\end{eqnarray}
and the {\em deviation function} $\Gamma(k_0, \hat{B})$ is
\begin{equation} 
\Gamma(k_0,\hat{B})= \cosh^2 (\frac{\pi}{2} \sqrt{4 \hat{B} e^{-x_0}-1})\,.
\end{equation}
When $B=0$ and $\Omega^{in} > \Omega^{out}$, then it is clear from
Eq. (\ref{betaka}) that the    
spectrum of created particles is nearly thermal to high accuracy,
\begin{equation}
\left|\frac{\beta_n}{\alpha_n}\right|^2 \simeq e^{-4 \pi \sqrt{C}} \,,
\end{equation} 
as expected in de Sitter expansion. However, when $B\neq 0$, $\gamma_0
\neq 0$, at the mode crossing time 
$n=H a(\eta)$, we can write: 
\begin{equation}
\left| \frac{\beta_n}{\alpha_n} \right|^2 \approx e^{-4 \pi \sqrt{C}} 
\left[\frac{1 + \frac{ \Gamma(k_0, B)}{ \sinh^2 2 \pi \hat{\Omega}_-}}
{1 + \frac{ \Gamma(k_0, B)}{ \sinh^2 2 \pi \hat{\Omega}_+}} \right] \,.
\end{equation}  
The expression in the squared bracket in the above equation contains
the deviation from scale invariance. The deviation $\Gamma(k_0,B)$ is 
larger at low values of the momentum modification scale, $x_0 \ll
1$. On the other hand, $\Gamma(k_0,B)$ is suppressed around large
scales, $x_0 \simeq 1$. The same results about the scale dependence of
the deviation function were obtained by using CMBFAST code
(Figs. \ref{fig2}).

\end{document}